\documentclass[12pt]{article}
\usepackage{amsfonts}

\begin{document}

\title{Algebraic Investigation of the Soft Seven Sphere}

\author{Khaled Abdel-Khalek \footnote{khaled@le.infn.it}\ \footnote{Address after 15 Feberuary 2000: 
Feza G\"ursey Institute, P.O. Box 6, 81220 \c{C}engelk\"oy, 
 \'Istanbul, Turkey}\\ Dipartimento di Fisica, Universit\`a di Lecce\\
- Lecce, 73100, Italy -
}
\date{February 2000}
\maketitle
\begin{abstract}
We investigate  the seven sphere as a soft Lie algebra i.e. an algebra with structure
functions instead of structure constants. We calculate its structure functions explicitly
and also discuss some relevant points such as the validity of the Jacobi identities. 
Furthermore, we emphasis some important features such as the pointwise reduction, closure
and some other consistency checks.
\end{abstract}

Ring division algebras play fundamental roles in mathematics from algebra to
geometry and topology with many different applications. The applicability of
real and complex numbers in physics is not in question. Quaternions which
may be represented as Pauli matrices are also important.

The use of octonions in physics is the problem. Non associative algebra
appeared for the first time in physics when Jordan, van Neuman and Wigner
introduced commutative but non-associative operators -- Jordan algebras --
for the construction of a new quantum mechanics\cite{s1}. More recently,
after the proposed eightfold way by Gell-Mann and Ne'eman, there were some
octonionic rivals for $SU\left( 3\right) $ such as $G_{2},\;SO\left(
7\right) ,\;SO\left( 8\right) $ and others\cite{s2}. Another serious step
was taken by G\"{u}naydin and G\"{u}rsey \cite{s3} when they present in
their work a systematic study of the octonionic algebraic structure. The
abandon of associativity means the existence of nonobservable states.
G\"{u}naydin and G\"{u}rsey used these nonobservable states to explain the
quark confinement phenomena.

Later on octonions entered the Grand Unified Theories era with different
applications. In \cite{s4}, G\"{u}rsey suggested that the exceptional group $%
F_{4}$ might be used to describe the internal charge space of particles.
Even the other exceptional groups $E_{6},E_{7},E_{8}$ have been utilized to
provide larger GUT models \cite{s5}. Starting from the eighties, new
applications of ring division algebras in physics were found. The instanton
problem \cite{s6}, supersymmetry \cite{s7}. Application of octonions to
supergravity spontaneous compactification was a very important and active
field of research during the mid eighties. Especially compactification of $%
d=11$ supergravity over $S^{7}$ to 4 dimensions. It is an impossible task to
list all the relevant papers, so we direct the interested reader to the
physics report \cite{s8} written by Duff, Nilsson and Pope where a lot of
references are given. We just mention that the first indication of the
octonionic nature of this problem appeared in the Englert solution of $d=11$
supergravity compactification over $S^{7}$\cite{s9} and a systematic study
along this line has been carried out in \cite{s10}\cite{s11}. The relations
between superstrings (p-branes) and octonions had been considered from many
different points of view, the reader may consult the references given in 
\cite{s12} for details.

Recently, a new line of attack has been proposed. Interpreting octonions as
a soft Lie algebra (algebra with strucuture functions instead of structure
constants \cite{s13}) proved to be useful for breaking the N = 8
superconformal barrier of models composed of spin 2 field, spin 3/2 fields,
fermions and spin one current. Englert, Sevrin, Troost, van Proeyen and
Spindel\cite{s14} constructed two different N = 8 superconformal algebras.
Latter this soft seven sphere (SSS) algebra was investigated by different
people. Berkovits \cite{s15} proved that the Green-Schwarz action admits
such algebra as an underlying symmetry, Cederwall and Preitschopf \cite{s16}
made a detailed study of this SSS algebra. Then they with Brink \cite{s17}
studied the light cone formulation of the Green-Schwarz action using such
SSS techniques. Also Samtleben \cite{s18} studied further the OPE structure
of the N = 8 superconformal algebra based on this SSS algebra.

In \cite{s13}, Sohnius used soft algebras with structure functions that vary
over space-time as well as over an internal gauge manifold. In this work we
use only soft algebras with structure functions that vary over the internal
gauge manifold (the fiber) not the base space-time manifold. In the first
section, we introduce the seven sphere as a soft algebra, we calculate its
structure functions explicitly and also discuss some relevant points such as
the validity of the Jacobi identity. Furthermore, we emphasis some important
features such as the pointwise reduction, closure, and some other
consistency checks. In the second section, we give more general
construction. We reformulate, in the third section, some standard Lie group
results putting them in a form suitable to subsequent applications.

\section{Octonions and the Soft Seven Sphere}

We start by recalling the non-associative octonion algebra. A generic
octonion number is 
\begin{equation}
\varphi =\varphi _{0}e_{0}+\varphi _{i}e_{i}=\varphi _{\mu }e_{\mu
}.\;\;\;\;\;\left[ i=1..7,\mu =0..7,\varphi _{\mu }\in \Bbb{R}\right]
\end{equation}
such that $e_{0}=1$ and the other seven imaginary units satisfy $%
e_{i}e_{j}=-\delta _{ij}+f_{ijk}e_{k}\Longleftrightarrow
[e_{i},e_{j}]=2f_{ijk}e_{k}$ where $f_{ijk}$ is completely antisymmetric and
equals one for any of the following three-cycles (123), (145), (246), (347),
(176), (257), (365) and the associator 
\begin{equation}
\lbrack e_{i},e_{j},e_{k}]=(e_{i}e_{j})e_{k}-e_{i}(e_{j}e_{k})\qquad ,
\end{equation}
is non-zero for any three elements that are not in the same three cycles and
is completely antisymmetric The following formula or any of its
generalization is thus ambiguous 
\begin{equation}
e_{1}e_{5}e_{7}=\left\{ 
\begin{array}{c}
(e_{1}e_{5})e_{7}=-e_{3} \\ 
\mbox{or} \\ 
e_{1}(e_{5}e_{7})=e_{3}
\end{array}
\right.  \label{hj1}
\end{equation}
so the best way is to define the action of the imaginary units in a certain
direction. In the spirit of Englert, Sevrin, Troost, Van Proeyen and Spindel%
\cite{s14} (also look at \cite{s19} and \cite{s20}) we define the left
action of octonionic operators $\delta _{i}$ by 
\begin{equation}
\delta _{i}\varphi =(e_{i}\varphi )
\end{equation}
implying that the following equation is well defined 
\begin{equation}
\delta _{i}\delta _{j}\varphi =\delta _{i}(\delta _{j}\varphi )=\delta
_{i}(e_{j}\varphi )=e_{i}(e_{j}\varphi ),
\end{equation}
then eq. (\ref{hj1}) reads unambiguously as 
\begin{equation}
\delta _{1}\delta _{5}e_{7}=(e_{1}(e_{5}e_{7}))=e_{3}.
\end{equation}
Since octonions are also non-commutative, we must also differentiate between
left and right action. Using the barred notation \cite{s21}, we introduce
right action as 
\begin{equation}
1|\delta _{i}~\varphi =(\varphi e_{i})\qquad ,
\end{equation}
for example 
\begin{equation}
(1|\delta _{i})(1|\delta _{j})\varphi =(1|\delta _{i})(1|\delta _{j}\varphi
)=1|\delta _{i}(\varphi e_{j})=((\varphi e_{j})e_{i})\qquad .
\end{equation}
As we shall see shortly, we can express the associator in terms of left and
right operators. The imaginary octonionic units generate the seven sphere $%
S^{7}$which has many properties similar to Lie algebras and/or Lie groups. $%
S^{7}$and Lie groups are the only non-flat compact parallelizable manifolds 
\cite{s22}.

The important point for evaluating any Lie algebra is the commutator, so
let's examine 
\begin{eqnarray}
\lbrack \delta _{i},\delta _{j}]\varphi &=&e_{i}(e_{j}\varphi
)-e_{j}(e_{i}\varphi )=2f_{ijk}e_{k}\varphi -2[e_{i},e_{j},\varphi ] 
\nonumber \\
&=&2f_{ijk}e_{k}\varphi +2[e_{i},\varphi ,e_{j}].
\end{eqnarray}
now the last term can be written as 
\begin{eqnarray}
\lbrack e_{i},\varphi ,e_{j}] &=&(e_{i}\varphi )e_{j}-e_{i}(\varphi
e_{j})=-\delta _{i}(1|\delta _{j})\varphi +(1|\delta _{j})\delta _{i}\varphi
\nonumber \\
&=&-[\delta _{i},1|\delta _{j}]\varphi ,
\end{eqnarray}
thus our commutator can be rewritten as 
\begin{equation}
\lbrack \delta _{i},\delta _{j}]\varphi =2f_{ijk}e_{k}\varphi
+2[e_{i},\varphi ,e_{j}]=2f_{ijk}\delta _{k}\varphi -2[\delta _{i},1|\delta
_{j}]\varphi .  \label{jk2}
\end{equation}
Note that right operators are necessary because the last term the associator
can never be written in terms of left operators alone.

After simple calculations, one concludes that the octonionic imaginary units
are determined completely by (\ref{jk2}) and the following equations 
\begin{eqnarray}
\left[ 1|\delta _{i},1|\delta _{j}\right] \varphi &=&-2f_{ijk}1|\delta
_{k}\varphi -2[\delta _{i},1|\delta _{j}]\varphi  \label{ll2} \\
\{\delta _{i},\delta _{j}\}\varphi &=&-2\delta _{ij}\varphi  \label{ll3} \\
\{1|\delta _{i},1|\delta _{j}\}\varphi &=&-2\delta _{ij}\varphi \qquad ,
\label{ll4}
\end{eqnarray}
where the $\delta _{ij}$ in (\ref{ll3}) and (\ref{ll4}) are the standard
Kronecker delta tensor.

It has been proved in \cite{s14}\cite{s16}, using three different ways, that
the $\delta _{i}$ algebra is associative. Thus a representation theory in
terms of matrices should be possible. Indeed, in \cite{s20}, we have derived
an algebra completely isomorphic to (\ref{jk2}--\ref{ll4}) by exploiting the
idea that octonions can be used as a basis for any $8\times 8$ real matrix.
we have two sets of matrices, essentially, 
\begin{equation}
\begin{array}{ccccc}
\delta _{i} & \Longleftrightarrow & (\Bbb{E}_{i})_{\mu \nu } & = & \delta
_{0\mu }\delta _{i\nu }-\delta _{0\nu }\delta _{i\mu }-f_{i\mu \nu }, \\ 
1|\delta _{i} & \Longleftrightarrow & (1|\Bbb{E}_{i})_{\mu \nu } & = & 
\delta _{0\mu }\delta _{i\nu }-\delta _{0\nu }\delta _{i\mu }+f_{i\mu \nu }.
\end{array}
\label{ee1}
\end{equation}
The set of matrices $\Bbb{E}_{i}$ and $1|\Bbb{E}_{i}\,\ $have appeared in
different octonionic works e.g. \cite{s3}\cite{s10}\cite{s23}. We suggest
that their most appropriate names should be \emph{the canonical left and
right octonionic structure at the north/south pole of the seven sphere}. By
explicit calculation, one finds that 
\begin{eqnarray}
\lbrack \Bbb{E}_{i},\Bbb{E}_{j}]\varphi &=&2f_{ijk}\Bbb{E}_{k}\varphi -2[%
\Bbb{E}_{i},1|\Bbb{E}_{j}]~\varphi  \label{lll2} \\
\left[ 1|\Bbb{E}_{i},1|\Bbb{E}_{j}\right] \varphi &=&-2f_{ijk}1|\Bbb{E}%
_{k}~\varphi -2[\Bbb{E}_{i},1|\Bbb{E}_{j}]~\varphi  \label{llll2} \\
\{\Bbb{E}_{i},\Bbb{E}_{j}\}\varphi &=&-2\delta _{ij}\varphi  \label{lll3} \\
\{1|\Bbb{E}_{i},1|\Bbb{E}_{j}\}\varphi &=&-2\delta _{ij}\varphi  \label{lll4}
\end{eqnarray}
where $\varphi $ is represented by a column matrix 
\[
\varphi ^{t}=\left( 
\begin{array}{llllllll}
\varphi _{0} & \varphi _{1} & \varphi _{2} & \varphi _{3} & \varphi _{4} & 
\varphi _{5} & \varphi _{6} & \varphi _{7}
\end{array}
\right) \quad 
\]
The word ``isomorphic'' above is justified since \{$\delta _{i}$\} is
associative \cite{s14}\cite{s16} and the same holds obviously for our \{$%
\Bbb{E}_{i}$ \} as they are written in terms of matrices. Our Jacobian
identities are 
\begin{equation}
\lbrack \delta _{i},[\delta _{j},\delta _{k}]]\varphi +[\delta _{j},[\delta
_{k},\delta _{i}]]\varphi +[\delta _{k},[\delta _{i},\delta _{j}]]\varphi
=0\qquad ,  \label{jac1}
\end{equation}
or 
\begin{equation}
\lbrack \Bbb{E}_{i},[\Bbb{E}_{j},\Bbb{E}_{k}]]\varphi +[\Bbb{E}_{j},[\Bbb{E}%
_{k},\Bbb{E}_{i}]]\varphi +[\Bbb{E}_{k},[\Bbb{E}_{i},\Bbb{E}_{j}]]\varphi
=0\qquad .  \label{jac2}
\end{equation}
We shall return to these identities again at the end of this section.

Fixing the direction of the application for any imaginary octonionic units
extracts a part of the algebra that respects associativity, but a certain
price has to be paid. The presence of the $\varphi $ is essential and either
of \{$\Bbb{E}_{i}$\} or \{$1|\Bbb{E}_{i}$\} is an open algebra, they don't
close upon the action of the commutator. Here comes the second step, the
soft Lie algebra idea. It is clear that the right hand side of (\ref{jk2})
or (\ref{lll2}) has a complicated $\varphi $ dependence. Knowing that the
seven sphere has a torsion that varies from one point to another \cite{s10}%
\cite{s22} and mimicking the Lie group case where the structure constants
are proportional to the fixed group torsion, it is natural to propose that (%
\ref{jk2}) may be redefined \cite{s14} as 
\begin{equation}
\lbrack \delta _{i},\delta _{j}]\varphi =2f_{ijk}^{\left( +\right) }(\varphi
)\delta _{k}\varphi \qquad ,
\end{equation}
where $f_{ijk}^{(+)}(\varphi )$ are structure functions that vary over the
whole $S^{7}$manifold. These structure functions $f_{ijk}^{(+)}(\varphi )$
were computed previously using different properties of the associator and
some other octonionic identities in \cite{s10}\cite{s14}\cite{s22}\cite{s16}%
. Here we use our matrix representation to give another alternative way to
calculate $f_{ijk}^{(+)}(\varphi )$%
\begin{equation}
\lbrack \Bbb{E}_{i},\Bbb{E}_{j}]{\ \varphi }=2f_{ijk}^{(+)}(\varphi )\Bbb{E}%
_{k}{\ \varphi }\qquad .  \label{tor1}
\end{equation}
Let's do it for the following example 
\begin{equation}
\lbrack \Bbb{E}_{1},\Bbb{E}_{2}]{\ \varphi }=2f_{12k}^{(+)}(\varphi )\Bbb{E}%
_{k}{\ \varphi }\qquad ,
\end{equation}
which is equivalent to the following eight equations , 
\begin{eqnarray}
&&
\begin{array}{l}
_{\varphi _{3}=\varphi _{1}f_{121}^{(+)}(\varphi )+\varphi
_{2}f_{122}^{(+)}(\varphi )+\varphi _{3}f_{123}^{(+)}(\varphi )+\varphi
_{4}f_{124}^{(+)}(\varphi )+\varphi _{5}f_{125}^{(+)}(\varphi )+\varphi
_{6}f_{126}^{(+)}(\varphi )+\varphi _{7}f_{127}^{(+)}(\varphi ),} \\ 
_{\varphi _{2}=\varphi _{2}f_{123}^{(+)}(\varphi )-\varphi
_{0}f_{121}^{(+)}(\varphi )-\varphi _{3}f_{122}^{(+)}(\varphi )-\varphi
_{5}f_{124}^{(+)}(\varphi )+\varphi _{4}f_{125}^{(+)}(\varphi )+\varphi
_{7}f_{126}^{(+)}(\varphi )-\varphi _{6}f_{127}^{(+)}(\varphi ),} \\ 
_{\varphi _{1}=\varphi _{0}f_{122}^{(+)}(\varphi )-\varphi
_{3}f_{121}^{(+)}(\varphi )+\varphi _{1}f_{123}^{(+)}(\varphi )+\varphi
_{6}f_{124}^{(+)}(\varphi )+\varphi _{7}f_{125}^{(+)}(\varphi )-\varphi
_{4}f_{126}^{(+)}(\varphi )-\varphi _{5}f_{127}^{(+)}(\varphi ),} \\ 
_{\varphi _{0}=\varphi _{2}f_{121}^{(+)}(\varphi )-\varphi
_{1}f_{122}^{(+)}(\varphi )+\varphi _{0}f_{123}^{(+)}(\varphi )+\varphi
_{7}f_{124}^{(+)}(\varphi )-\varphi _{6}f_{125}^{(+)}(\varphi )+\varphi
_{5}f_{126}^{(+)}(\varphi )-\varphi _{4}f_{127}^{(+)}(\varphi ),} \\ 
_{\varphi _{7}=\varphi _{0}f_{124}^{(+)}(\varphi )-\varphi
_{5}f_{121}^{(+)}(\varphi )-\varphi _{6}f_{122}^{(+)}(\varphi )-\varphi
_{7}f_{123}^{(+)}(\varphi )+\varphi _{1}f_{125}^{(+)}(\varphi )+\varphi
_{2}f_{126}^{(+)}(\varphi )+\varphi _{3}f_{127}^{(+)}(\varphi ),} \\ 
_{\varphi _{6}=\varphi _{7}f_{122}^{(+)}(\varphi )-\varphi
_{4}f_{121}^{(+)}(\varphi )-\varphi _{6}f_{123}^{(+)}(\varphi )+\varphi
_{1}f_{124}^{(+)}(\varphi )-\varphi _{0}f_{125}^{(+)}(\varphi )+\varphi
_{3}f_{126}^{(+)}(\varphi )-\varphi _{2}f_{127}^{(+)}(\varphi ),} \\ 
_{\varphi _{5}=\varphi _{7}f_{121}^{(+)}(\varphi )+\varphi
_{4}f_{122}^{(+)}(\varphi )-\varphi _{5}f_{123}^{(+)}(\varphi )-\varphi
_{2}f_{124}^{(+)}(\varphi )+\varphi _{3}f_{125}^{(+)}(\varphi )+\varphi
_{0}f_{126}^{(+)}(\varphi )-\varphi _{1}f_{127}^{(+)}(\varphi ),} \\ 
_{\varphi _{4}=\varphi _{6}f_{121}^{(+)}(\varphi )-\varphi
_{5}f_{122}^{(+)}(\varphi )-\varphi _{4}f_{123}^{(+)}(\varphi )+\varphi
_{3}f_{124}^{(+)}(\varphi )+\varphi _{2}f_{125}^{(+)}(\varphi )-\varphi
_{1}f_{126}^{(+)}(\varphi )-\varphi _{0}f_{127}^{(+)}(\varphi ).}
\end{array}
\nonumber \\
&&  \label{eqns}
\end{eqnarray}
We now solve these equations for the seven unknown $f_{12i}^{(+)}\left(
\varphi \right) $. \ We find 
\[
\begin{array}{l}
f_{121}^{(+)}\left( \varphi \right) =f_{122}^{(+)}\left( \varphi \right) =0,
\\ 
f_{124}^{(+)}\left( \varphi \right) =+2{\frac{{\varphi _{0}}{\varphi _{7}}-{%
\varphi _{5}}{\varphi _{2}}+{\varphi _{6}}{\varphi _{1}}+{\varphi _{3}}{%
\varphi _{4}}}{r^{2}},} \\ 
f_{125}^{(+)}\left( \varphi \right) =-2{\frac{{\varphi _{0}}{\varphi _{6}}-{%
\varphi _{3}}{\varphi _{5}}-{\varphi _{1}}{\varphi _{7}}-{\varphi _{2}}{%
\varphi _{4}}}{r^{2}},} \\ 
f_{126}^{(+)}\left( \varphi \right) =+2{\frac{{\varphi _{0}}{\varphi _{5}}-{%
\varphi _{1}}{\varphi _{4}}+{\varphi _{7}}{\varphi _{2}}+{\varphi _{3}}{%
\varphi _{6}}}{r^{2}},} \\ 
f_{127}^{(+)}\left( \varphi \right) =-2{\frac{{\varphi _{0}}{\varphi _{4}}+{%
\varphi _{6}}{\varphi _{2}}+{\varphi _{1}}{\varphi _{5}}-{\varphi _{3}}{%
\varphi _{7}}}{r^{2}},} \\ 
f_{123}^{(+)}\left( \varphi \right) ={\frac{{\varphi _{0}}^{2}-{\varphi _{6}}%
^{2}-{\varphi _{5}}^{2}+{\varphi _{2}}^{2}-{\varphi _{4}}^{2}+{\varphi _{1}}%
^{2}+{\varphi _{3}}^{2}-{\varphi _{7}}^{2}}{r^{2}},}
\end{array}
\]
where 
\[
r^{2}=(\varphi _{0}^{2}{\ +}\varphi _{1}^{2}{\ +}\varphi _{2}^{2}{\ +}%
\varphi _{3}^{2}{\ +}\varphi _{4}^{2}{\ +}\varphi _{5}^{2}{\ +}\varphi
_{6}^{2}{\ +}\varphi _{7}^{2}).
\]
Along the same lines we can calculate all the structure functions, we give
all of them in Appendix A. What we have just calculated is commonly called
the (+) torsion\cite{s10}, we can find the (--) torsion by replacing the
left by right multiplication in (\ref{tor1}) 
\begin{equation}
\lbrack 1|\Bbb{E}_{i},1|\Bbb{E}_{j}]{\ \varphi }=2f_{ijk}^{(-)}(\varphi )1|%
\Bbb{E}_{k}{\ \varphi }.
\end{equation}
we find 
\[
\begin{array}{l}
f_{121}^{(-)}\left( \varphi \right) =f_{122}^{(-)}\left( \varphi \right) =0,
\\ 
f_{124}^{(-)}\left( \varphi \right) =+2{\frac{{\varphi _{0}}{\varphi _{7}}+{%
\varphi _{5}}{\varphi _{2}}-{\varphi _{6}}{\varphi _{1}}-{\varphi _{3}}\ ,{%
\varphi _{4}}}{r^{2}},} \\ 
f_{125}^{(-)}\left( \varphi \right) =-2{\frac{{\varphi _{0}}{\varphi _{6}}+{%
\varphi _{3}}{\varphi _{5}}+{\varphi _{1}}{\varphi _{7}}+{\varphi _{2}}{%
\varphi _{4}}}{r^{2}},} \\ 
f_{126}^{(-)}\left( \varphi \right) =+2{\frac{{\varphi _{0}}{\varphi _{5}}+{%
\varphi _{1}}{\varphi _{4}}-{\varphi _{7}}{\varphi _{2}}-{\varphi _{3}}\ ,{%
\varphi _{6}}}{r^{2}},} \\ 
f_{127}^{(-)}\left( \varphi \right) =-2{\frac{{\varphi _{0}}{\varphi _{4}}-{%
\varphi _{6}}{\varphi _{2}}-{\varphi _{1}}{\varphi _{5}}+{\varphi _{3}}{%
\varphi _{7}}}{r^{2}},} \\ 
f_{123}^{(-)}\left( \varphi \right) =-{\frac{{\varphi _{0}}^{2}-{\varphi _{6}%
}^{2}-{\varphi _{4}}^{2}+{\varphi _{2}}^{2}+{\varphi _{1}}^{2}-{\varphi _{7}}%
^{2}-{\varphi _{5}}^{2}+{\varphi _{3}}^{2}}{r^{2}},}
\end{array}
\]
the remaining $f_{ijk}^{(-)}\left( \varphi \right) $ are listed in appendix
A.

Let's pause for a moment and note some of the evident features of these $%
f_{ijk}^{\left( \pm \right) }(\varphi )$,

\begin{itemize}
\item  One notices immediately that at $\varphi ^{t}$=(1,0,0,0,0,0,0,0) /
(-1,0,0,0,0,0,0,0), the north / south pole (NP/SP), we recover the
octonionic structure constants: $%
f_{ijk}^{(+)}(NP/SP)=-f_{ijk}^{(-)}(NP/SP)=f_{ijk}$ and any non-standard
cycles vanishes e.g. $f_{567}^{\left( \pm \right) }(NP/SP)=0$.

\item  Our construction started from a given multiplication table and as
there are different choices \cite{s24}, we can have different families.

\item  Over S$^{7}$, $\partial _{\mbox{S}^{7}}f_{ijk}^{\left( \pm \right)
}(\varphi )\neq 0.$ This is a very important characteristic of the seven
sphere.
\end{itemize}

To manifest the $\varphi $ dependence, let's give a non--tivial example. To
simplify the notations, we use here $(i,j,k)^{\left( \pm \right) }$ for $%
f_{ijk}^{\left( \pm \right) }(\varphi ).$ At $\left( \varphi _{\mu }=\frac{%
\mu +1}{\sqrt{204}}\right) $, we find 
\begin{eqnarray}
&& 
\begin{array}{c}
\begin{array}{lll}
(1,2,3)^{(+)}=-12/17, & (2,5,7)^{(+)}=4/51, & (1,5,6)^{(+)}=1/51, \\ 
(1,4,5)^{(+)}=-6/17, & (1,7,6)^{(+)}=8/51, & (3,6,5)^{(+)}=0, \\ 
(4,3,7)^{(+)}=-2/51, & (4,2,6)^{(+)}=3/17, & 
\end{array}
\\ 
\begin{array}{lll}
(1,2,4)^{(+)}=4/17, & (1,5,2)^{(+)}=-8/17, & (3,5,4)^{(+)}=-44/51, \\ 
(5,6,7)^{(+)}=-10/17, & (1,3,4)^{(+)}=-5/17, & (4,1,6)^{(+)}=-14/17, \\ 
(1,5,7)^{(+)}=-40/51, & (3,5,7)^{(+)}=-2/17, & (3,1,6)^{(+)}=-14/51, \\ 
(2,3,5)^{(+)}=-23/51, & (1,7,4)^{(+)}=-4/17, & (4,5,6)^{(+)}=-16/51, \\ 
(2,6,5)^{(+)}=-38/51, & (1,6,2)^{(+)}=-8/17, & (6,3,2)^{(+)}=-22/51, \\ 
(1,3,5)^{(+)}=-10/51, & (2,4,3)^{(+)}=-2/17, & (4,3,6)^{(+)}=-20/51, \\ 
(3,7,6)^{(+)}=-13/17, & (1,2,7)^{(+)}=-1/17, & (4,2,7)^{(+)}=-16/17, \\ 
(2,7,3)^{(+)}=-16/17, & (2,6,7)^{(+)}=-4/51, & (7,1,3)^{(+)}=-28/51, \\ 
(2,4,5)^{(+)}=2/17, & (4,7,5)^{(+)}=-7/51, & (4,6,7)^{(+)}=-10/51.
\end{array}
\end{array}
\nonumber \\
&&  \label{xxx1}
\end{eqnarray}
We have some kind of dynamical Lie algebra of seven generators with
structure ``constants'' that change their values from one point to another.
Let us emphasis the difference between considering $\Bbb{E}_{i}$ as an open
algebra or as elements of a soft seven sphere, observe that 
\begin{equation}
\begin{array}{ll}
\left[ \Bbb{E}_{1},\Bbb{E}_{2}\right] & =2\Bbb{E}_{3}-2\left[ \Bbb{E}_{1},1|%
\Bbb{E}_{2}\right] ,
\end{array}
\end{equation}
but 
\begin{equation}
\begin{array}{lll}
\left[ \Bbb{E}_{1},\Bbb{E}_{2}\right] \Phi & =2f_{123}^{\left( +\right)
}\left( \varphi \right) \Bbb{E}_{3}\Phi & +2f_{124}^{\left( +\right) }\left(
\varphi \right) \Bbb{E}_{4}\Phi \\ 
& +2f_{125}^{\left( +\right) }\left( \varphi \right) \Bbb{E}_{5}\Phi & 
+2f_{126}^{\left( +\right) }\left( \varphi \right) \Bbb{E}_{6}\Phi
+2f_{127}^{\left( +\right) }\left( \varphi \right) \Bbb{E}_{7}\Phi .
\end{array}
\label{soft}
\end{equation}
At the NP 
\begin{equation}
\Phi _{NP}^{t}=\left( 
\begin{array}{llllllll}
1 & 0 & 0 & 0 & 0 & 0 & 0 & 0
\end{array}
\right)
\end{equation}
we still have 
\begin{equation}
\left[ \Bbb{E}_{1},\Bbb{E}_{2}\right] =2\Bbb{E}_{3}-2\left[ \Bbb{E}_{1},1|%
\Bbb{E}_{2}\right]
\end{equation}
whereas 
\begin{eqnarray*}
\left[ \Bbb{E}_{1},\Bbb{E}_{2}\right] \Phi _{NP} &=&2f_{12k}^{\left(
+\right) }(\varphi _{NP})\Bbb{E}_{k}\Phi _{NP} \\
&=&2\Bbb{E}_{3}\Phi _{NP}
\end{eqnarray*}
\begin{eqnarray*}
&&\left( 
\begin{array}{llllllll}
0 & 0 & 0 & -2 & 0 & 0 & 0 & 0 \\ 
0 & 0 & -2 & 0 & 0 & 0 & 0 & 0 \\ 
0 & 2 & 0 & 0 & 0 & 0 & 0 & 0 \\ 
2 & 0 & 0 & 0 & 0 & 0 & 0 & 0 \\ 
0 & 0 & 0 & 0 & 0 & 0 & 0 & 2 \\ 
0 & 0 & 0 & 0 & 0 & 0 & -2 & 0 \\ 
0 & 0 & 0 & 0 & 0 & 2 & 0 & 0 \\ 
0 & 0 & 0 & 0 & -2 & 0 & 0 & 0
\end{array}
\right) \left( 
\begin{array}{l}
1 \\ 
0 \\ 
0 \\ 
0 \\ 
0 \\ 
0 \\ 
0 \\ 
0
\end{array}
\right) =\left( 
\begin{array}{l}
0 \\ 
0 \\ 
0 \\ 
2 \\ 
0 \\ 
0 \\ 
0 \\ 
0
\end{array}
\right) \\
&=&\left( 
\begin{array}{llllllll}
0 & 0 & 0 & -2 & 0 & 0 & 0 & 0 \\ 
0 & 0 & -2 & 0 & 0 & 0 & 0 & 0 \\ 
0 & 2 & 0 & 0 & 0 & 0 & 0 & 0 \\ 
2 & 0 & 0 & 0 & 0 & 0 & 0 & 0 \\ 
0 & 0 & 0 & 0 & 0 & 0 & 0 & -2 \\ 
0 & 0 & 0 & 0 & 0 & 0 & 2 & 0 \\ 
0 & 0 & 0 & 0 & 0 & -2 & 0 & 0 \\ 
0 & 0 & 0 & 0 & 2 & 0 & 0 & 0
\end{array}
\right) \left( 
\begin{array}{l}
1 \\ 
0 \\ 
0 \\ 
0 \\ 
0 \\ 
0 \\ 
0 \\ 
0
\end{array}
\right) .
\end{eqnarray*}

At $\Phi _{W}\equiv \left( \varphi _{\mu }=\frac{\mu +1}{\sqrt{204}}\right)
, $ we still have 
\[
\left[ \Bbb{E}_{1},\Bbb{E}_{2}\right] =2\Bbb{E}_{3}-2\left[ \Bbb{E}_{1},1|%
\Bbb{E}_{2}\right] 
\]
we find that 
\begin{eqnarray*}
\left[ \Bbb{E}_{1},\Bbb{E}_{2}\right] \Phi _{W} &=&2f_{12k}^{\left( +\right)
}(\varphi _{W})\Bbb{E}_{k}\Phi _{W} \\
&=&\left( -\frac{24}{17}\Bbb{E}_{3}+\frac{8}{17}\Bbb{E}_{4}+\frac{16}{17}%
\Bbb{E}_{5}+\frac{16}{17}\Bbb{E}_{6}-\frac{2}{17}\Bbb{E}_{7}\right) \Phi _{W}
\end{eqnarray*}
\begin{eqnarray*}
&&\left( 
\begin{array}{llllllll}
0 & 0 & 0 & -2 & 0 & 0 & 0 & 0 \\ 
0 & 0 & -2 & 0 & 0 & 0 & 0 & 0 \\ 
0 & 2 & 0 & 0 & 0 & 0 & 0 & 0 \\ 
2 & 0 & 0 & 0 & 0 & 0 & 0 & 0 \\ 
0 & 0 & 0 & 0 & 0 & 0 & 0 & 2 \\ 
0 & 0 & 0 & 0 & 0 & 0 & -2 & 0 \\ 
0 & 0 & 0 & 0 & 0 & 2 & 0 & 0 \\ 
0 & 0 & 0 & 0 & -2 & 0 & 0 & 0
\end{array}
\right) \left( 
\begin{array}{l}
\frac{1}{\sqrt{204}} \\ 
\frac{2}{\sqrt{204}} \\ 
\frac{3}{\sqrt{204}} \\ 
\frac{4}{\sqrt{204}} \\ 
\frac{5}{\sqrt{204}} \\ 
\frac{6}{\sqrt{204}} \\ 
\frac{7}{\sqrt{204}} \\ 
\frac{8}{\sqrt{204}}
\end{array}
\right) =\left( 
\begin{array}{l}
\frac{-4}{51}\sqrt{51} \\ 
\frac{-1}{17}\sqrt{51} \\ 
\frac{2}{51}\sqrt{51} \\ 
\frac{1}{51}\sqrt{51} \\ 
\frac{8}{51}\sqrt{51} \\ 
\frac{-7}{51}\sqrt{51} \\ 
\frac{2}{17}\sqrt{51} \\ 
\frac{-5}{51}\sqrt{51}
\end{array}
\right) \\
&=&\left( 
\begin{array}{llllllll}
0 & 0 & 0 & \frac{24}{17} & \frac{-8}{17} & \frac{-16}{17} & \frac{-16}{17}
& \frac{2}{17} \\ 
0 & 0 & \frac{24}{17} & 0 & \frac{-16}{17} & \frac{8}{17} & \frac{-2}{17} & 
\frac{-16}{17} \\ 
0 & \frac{-24}{17} & 0 & 0 & \frac{-16}{17} & \frac{2}{17} & \frac{8}{17} & 
\frac{16}{17} \\ 
\frac{-24}{17} & 0 & 0 & 0 & \frac{2}{17} & \frac{16}{17} & \frac{-16}{17} & 
\frac{8}{17} \\ 
\frac{8}{17} & \frac{16}{17} & \frac{16}{17} & \frac{-2}{17} & 0 & 0 & 0 & 
\frac{24}{17} \\ 
\frac{16}{17} & \frac{-8}{17} & \frac{-2}{17} & \frac{-16}{17} & 0 & 0 & 
\frac{-24}{17} & 0 \\ 
\frac{16}{17} & \frac{2}{17} & \frac{-8}{17} & \frac{16}{17} & 0 & \frac{24}{%
17} & 0 & 0 \\ 
\frac{-2}{17} & \frac{16}{17} & \frac{-16}{17} & \frac{-8}{17} & \frac{-24}{%
17} & 0 & 0 & 0
\end{array}
\right) \left( 
\begin{array}{l}
\frac{1}{\sqrt{204}} \\ 
\frac{2}{\sqrt{204}} \\ 
\frac{3}{\sqrt{204}} \\ 
\frac{4}{\sqrt{204}} \\ 
\frac{5}{\sqrt{204}} \\ 
\frac{6}{\sqrt{204}} \\ 
\frac{7}{\sqrt{204}} \\ 
\frac{8}{\sqrt{204}}
\end{array}
\right) .
\end{eqnarray*}

We are not making a projection but a reformulation of the algebra. This fact
should always be kept in mind. The same happens in a non-trivial way for the
Jacobi identity, i.e. 
\begin{equation}
\left( f_{ijm}(\varphi )f_{mkt}(\varphi )+f_{jkm}(\varphi )f_{mit}(\varphi
)+f_{kim}(\varphi )f_{mjt}(\varphi )\right) \Bbb{E}_{t}\varphi =0,
\end{equation}
but, in general, 
\begin{equation}
\left( f_{ijm}(\varphi )f_{mkt}(\varphi )+f_{jkm}(\varphi )f_{mit}(\varphi
)+f_{kim}(\varphi )f_{mjt}(\varphi )\right) \neq 0.
\end{equation}
Another important feature is 
\begin{equation}
\left[ \Bbb{E}_{i},1|\Bbb{E}_{j}\right] =-2f_{ijk}\left( \varphi \right) 
\Bbb{E}_{k}+\left[ \Bbb{E}_{i},\Bbb{E}_{j}\right] =-2f_{ijk}\left( \varphi
\right) 1|\Bbb{E}_{k}+\left[ 1|\Bbb{E}_{i},1|\Bbb{E}_{j}\right]
\end{equation}
which is equal to zero iff $i=j$ but for the soft seven sphere, $\left[ \Bbb{%
E}_{i},1|\Bbb{E}_{j}\right] \varphi =0$ not only for $i=j$ but also at the
NP/SP for any i,j.

Lastly over any group manifold the left torsion equals minus the right
torsion, but for $S^{7}$ this is not in general true.

\section{More General Solutions}

In the previous section , we have used brute force to calculate $%
f_{ijk}^{\left( +\right) }\left( \varphi \right) $. There is another way,
smarter and easier. We have the following situation 
\begin{equation}
\Bbb{E}_{i}\Bbb{E}_{j}\;\varphi =\left( -\delta _{ij}+f_{ijk}^{\left(
+\right) }\left( \varphi \right) \Bbb{E}_{k}\right) \varphi
\end{equation}
but one can check that our $\Bbb{E}_{i}$ defines what Cartan calls pure
spinors \cite{s25}, 
\begin{equation}
\varphi ^{t}\Bbb{E}_{i}\varphi =0
\end{equation}
thus 
\begin{equation}
\varphi ^{t}\left( \Bbb{E}_{i}\Bbb{E}_{j}\right) \varphi =\varphi ^{t}\left(
-\delta _{ij}\right) \varphi ,
\end{equation}
using 
\begin{equation}
\left( \Bbb{E}_{k}\right) ^{-1}=-\Bbb{E}_{k}
\end{equation}
we find 
\begin{equation}
\varphi ^{t}\left( -\Bbb{E}_{k}\Bbb{E}_{i}\Bbb{E}_{j}\right) \varphi
=\varphi ^{t}\left( f_{ijk}^{\left( +\right) }\left( \varphi \right) \right)
\varphi
\end{equation}
but 
\begin{equation}
\varphi ^{t}\varphi =r^{2}
\end{equation}
which gives us 
\begin{equation}
f_{ijk}^{\left( +\right) }\left( \varphi \right) =\frac{\varphi ^{t}\left( -%
\Bbb{E}_{k}\Bbb{E}_{i}\Bbb{E}_{j}\right) \varphi }{r^{2}}.
\end{equation}
and 
\begin{equation}
f_{ijk}^{\left( -\right) }\left( \varphi \right) =\frac{\varphi ^{t}\left(
-1|\Bbb{E}_{k}\;\;1|\Bbb{E}_{i}\;\;1|\Bbb{E}_{j}\right) \varphi }{r^{2}}.
\end{equation}
There is another interesting property to note 
\begin{equation}
\varphi ^{t}\left[ \Bbb{E}_{i},1|\Bbb{E}_{j}\right] \varphi =0
\end{equation}
which may be the generalization of the standard Lie algebra relation, left
and right action commute everywhere over the group manifold.

The left and right torsions that we have constructed are not the only
parallelizable torsions of S$^{7}$. Our $\Bbb{E}_{i}$ and $1|\Bbb{E}_{i}$
are given in terms of the octonionic structure constants (\ref{ee1}) i.e.
the torsion at NP/SP. Considering two new points, we may define new sets of $%
\Bbb{E}_{i}$ and $1|\Bbb{E}_{i}$. As S$^{7}$ contains an infinity of points,
practically, we have an infinity of parallelizable torsion. If our method is
self contained and sufficient, we should be able to construct these infinity
of pointwise structures. Indeed, $\Bbb{E}_{i}\left( \varphi \right) $ and $1|%
\Bbb{E}_{i}\left( \varphi \right) $ are in general 
\begin{equation}
\begin{array}{ccccc}
\delta _{i} & \Longleftrightarrow & (\Bbb{E}_{i}(\varphi ))_{\mu \nu } & = & 
\delta _{0\mu }\delta _{i\nu }-\delta _{0\nu }\delta _{i\mu }-f_{i\mu \nu
}^{\left( +\right) }(\varphi ), \\ 
1|\delta _{i} & \Longleftrightarrow & (1|\Bbb{E}_{i}(\varphi ))_{\mu \nu } & 
= & \delta _{0\mu }\delta _{i\nu }-\delta _{0\nu }\delta _{i\mu }+f_{i\mu
\nu }^{\left( -\right) }(\varphi ),
\end{array}
\label{octon}
\end{equation}
in complete analogy with (\ref{jk2},\ref{ll2},\ref{ll3},\ref{ll4}). Of
course the soft Algebra idea should hold here as well as for the special $%
\left( \Bbb{E}_{i},1|\Bbb{E}_{i}\right) $ constructed in terms of the north
pole torsion. Repeating the calculation in terms of $\left( \Bbb{E}%
_{i}(\varphi ),1|\Bbb{E}_{i}(\varphi )\right) $. Let us introduce a new
vector field $\lambda $, 
\begin{equation}
\lambda ^{t}=\left( 
\begin{array}{llllllll}
\lambda _{0} & \lambda _{1} & \lambda _{2} & \lambda _{3} & \lambda _{4} & 
\lambda _{5} & \lambda _{6} & \lambda _{7}
\end{array}
\right) .
\end{equation}
We define two new generalized structure functions 
\begin{eqnarray}
\left[ \Bbb{E}_{i}(\varphi ),\Bbb{E}_{j}(\varphi )\right] \lambda
&=&2f_{ijk}^{(++)}(\varphi ,\lambda )\Bbb{E}_{k}(\varphi )\lambda \\
\left[ 1|\Bbb{E}_{i}(\varphi ),1|\Bbb{E}_{j}(\varphi )\right] \lambda
&=&2f_{ijk}^{(-\;-)}(\varphi ,\lambda )1|\Bbb{E}_{k}(\varphi )\lambda
\end{eqnarray}
where $f_{ijk}^{\left( \pm \;\pm \right) }\left( \varphi ,\lambda \right) $
have a very complicated structure, 
\begin{eqnarray}
f_{ijk}^{\left( ++\right) }\left( \varphi ,\lambda \right) &=&\frac{\lambda
^{t}\left( -\Bbb{E}_{k}\left( \varphi \right) \Bbb{E}_{i}\left( \varphi
\right) \Bbb{E}_{j}\left( \varphi \right) \right) \lambda }{r^{2}}, \\
f_{ijk}^{\left( -\;-\right) }\left( \varphi ,\lambda \right) &=&\frac{%
\lambda ^{t}\left( -1|\Bbb{E}_{k}\left( \varphi \right) \;\;1|\Bbb{E}%
_{i}\left( \varphi \right) \;\;1|\Bbb{E}_{j}\left( \varphi \right) \right)
\lambda }{r^{2}}
\end{eqnarray}

\section{Some Group Theory}

An arbitrary octonion can be associated to $\Bbb{R}^{8}=\Bbb{R}\oplus \Bbb{R}%
^{7}$ \cite{s26} where $\Bbb{R}$ denotes the subspace spanned by the
identity $e_{0}=1$. Octonions with unit length define the octonionic unit
sphere $S^{7}$. The isometries of octonions is described by $O(8)$ which may
be decomposed as 
\begin{equation}
O(8):\quad \quad H\oplus K\oplus E
\end{equation}
where $H$ is the 14 parameters $G_{2}$ algebra of the automorphism group of $%
octonions,$ K is the torsionful seven sphere $SO(7)/G_{2}$ and our E is the
round seven sphere $SO(8)/SO(7)$. In fact the different three non-equivalent
representation of O(8) - the vectorial so(8) and the two different spinorial 
$spin^{L}(8)$ and $spin^{R}(8)$, which are related by triality, can be
realized by suitable left and right octonionic multiplication. The reduction
of O(8) to O(7) induces $so(8)\longrightarrow so(7)\oplus 1$, $%
spin^{R}(8)\longrightarrow spin(7)$ and $spin^{L}(8)\longrightarrow spin(7)$.

We would like to show how to generate these different Lie algebras entirely
from our canonical left/right octonionic structures. We start from the $%
8\times 8$ gamma matrices $\gamma _{\mu \nu }^{i}$ in seven dimensions,
using $\delta _{ij}$ as our flat Euclidean metric, 
\begin{equation}
\{\gamma ^{i},\gamma ^{j}\}=2\delta ^{ij}\mathbf{1}_{8}\qquad ,
\end{equation}
where $i,j,\ldots =1,2,\ldots 7$ and $\mu ,\nu ,\ldots =0,1,2,\ldots 7$. We
can use either of the following choices 
\begin{equation}
\gamma _{+}^{j}=i\Bbb{E}_{j}\quad or\quad \gamma _{-}^{j}=i1|\Bbb{E}%
_{j}\qquad ,  \label{cliffo}
\end{equation}
of course the $i$ in the right hand sides is the imaginary complex unit.
This relates our antisymmetric, Hermitian and \ hence purely imaginary gamma
matrices to the canonical octonionic left/right structures. The
antisymmetric product of two gamma matrices will be denoted by 
\begin{equation}
\gamma ^{ij}=\gamma ^{[i}\gamma ^{j]}\qquad ,
\end{equation}
and we have \footnote{%
It is interesting to note that this equation may be used as an alternative
definition for the octonionic multiplication table.} 
\begin{equation}
\gamma ^{i}\gamma ^{j}\gamma ^{k}= \frac{1}{4!} \epsilon ^{ijklmnp}\gamma ^{l}\gamma
^{m}\gamma ^{n}\gamma ^{p}\qquad .
\end{equation}

The matrices $\gamma ^{ij}$ span the 21 generators $J^{ij}$of spin(7) in its
eight-dimensional spinor representation. The spinorial representation of
spin(7) can be enlarged to the left/right handed spinor representation of
spin(8) by different ways. The easiest one is to include either of $\pm \Bbb{%
E}_{i}$ or $\pm 1|\Bbb{E}_{i}$ \cite{s3}\cite{s23} defining $J^{i}=J^{i0}$,
so(8) can be written as 
\begin{eqnarray}
\lbrack J^{i},J^{i}] &=&2J^{ij} \\
\lbrack J^{i},J^{mn}] &=&2\delta ^{im}J^{n}-2\delta ^{in}J^{m} \\
\lbrack J^{ij},J^{kl}] &=&2\delta ^{jk}J^{il}+2\delta ^{il}J^{jk}-2\delta
^{ik}J^{jl}-2\delta ^{jl}J^{ik}.
\end{eqnarray}

The automorphism group of octonions is $G_{2}\subset SO(7)\subset SO(8)$. A
suitable basis for $G_{2}$ is \cite{s3}\cite{s23}\cite{s26} 
\begin{equation}
H_{ij}=f_{ijk}\left( \Bbb{E}_{k}-1|\Bbb{E}_{k}\right) -\frac{3}{2}\left[ 
\Bbb{E}_{i},1|\Bbb{E}_{j}\right] \qquad ,
\end{equation}
which implies the linear relations 
\begin{equation}
f_{ijk}H_{jk}=0\qquad ,
\end{equation}
These constraints enforce $H_{ij}$ to generate the 14 dimensional vector
space of $G_{2}$. There are different ways to represent the remaining seven
generators, denoted here by K, 
\begin{eqnarray}
\frac{so(7)}{G_{2}} &:&K_{v}^{\pm i}=\pm \frac{1}{2}\left( \Bbb{E}_{i}-1|%
\Bbb{E}_{i}\right) \qquad ,  \label{lo1} \\
\frac{spin(7)}{G_{2}} &:&K_{s}^{\pm i}=\pm \left( \frac{1}{2}\Bbb{E}_{i}+1|%
\Bbb{E}_{i}\right) \qquad ,  \label{lo2} \\
\frac{\overline{spin(7)}}{G_{2}} &:&{\overline{K}}_{s}^{\ \pm i}=\mp \left( 
\Bbb{E}_{i}+\frac{1}{2}1|\Bbb{E}_{i}\right) \qquad ,  \label{lo3}
\end{eqnarray}
Defining the conjugate representation\footnote{%
n.b. this definition is not matrix conjugation.} \ by 
\begin{equation}
{\overline{\Bbb{E}}}=-1|\Bbb{E}\qquad \mbox{and\qquad }{\overline{1|\Bbb{E}}}%
=-\Bbb{E}\qquad ,
\end{equation}
(\ref{lo1}) is self-conjugate while (\ref{lo2}) is octonionic-conjugate to (%
\ref{lo3}). The vector representation so(7) generated by $H_{ij}\oplus
K_{v}^{\pm i}$ is seven dimensional because $K_{v}^{\pm i}e_{0}=0$ whereas
the spin(7) representation generated by $H_{ij}\oplus K_{s}^{\pm i}$ is
eight dimensional.

To make apparent the role of the automorphism group $G_{2}$, the different
commutators of $\Bbb{E}$ and $1|\Bbb{E}$ may be written as 
\begin{eqnarray}
\left[ \Bbb{E}_{i},\Bbb{E}_{j}\right] &=&\frac{1}{3}\left( 4H_{ij}+2f_{ijk}%
\Bbb{E}_{k}+4f_{ijk}1|\Bbb{E}_{k}\right) \qquad , \\
\left[ 1|\Bbb{E}_{i},1|\Bbb{E}_{j}\right] &=&\frac{1}{3}\left(
4H_{ij}-4f_{ijk}\Bbb{E}_{k}-2f_{ijk}1|\Bbb{E}_{k}\right) \qquad , \\
\left[ \Bbb{E}_{i},1|\Bbb{E}_{j}\right] &=&\frac{1}{3}\left(
-2H_{ij}+2f_{ijk}\Bbb{E}_{k}-2f_{ijk}1|\Bbb{E}_{k}\right) \qquad .
\end{eqnarray}
or $G_{2}$ given by 
\begin{equation}
H_{ij}=\frac{1}{2}\left( [\Bbb{E}_{i},\Bbb{E}_{j}]+[1|\Bbb{E}_{i},1|\Bbb{E}%
_{j}]+[\Bbb{E}_{i},1|\Bbb{E}_{j}]\right) \qquad .
\end{equation}
Thus as we promised, the $\Bbb{E}$ and $1|\Bbb{E}$ are the necessary and the
sufficient building blocks for expressing the different Lie algebras and
coset representations related to the seven sphere. Note that all the
constructions given in this section start from the Clifford algebra relation
(\ref{cliffo}), and the formulation holds equally for $\Bbb{E}\left( \varphi
\right) $ and $1|\Bbb{E}\left( \varphi \right) $.

I am grateful to C. Imbimbo, P. Rotelli, G. Thompson, A. Van Proeyen for useful comments.
Also, I would like to thank S. Marchiafava and F. Englert for encouragements.

\appendix

\section{The structure function of the soft seven sphere}

The seven standard cycles are given by 
\[
\begin{array}{l}
{{f}_{123}^{(+)}{(\varphi )}}={{-f}_{123}^{(-)}{(\varphi )}}={\frac{{\varphi
_{0}}^{2}-{\varphi _{6}}^{2}-{\varphi _{5}}^{2}+{\varphi _{2}}^{2}-{\varphi
_{4}}^{2}+{\varphi _{1}}^{2}+{\varphi _{3}}^{2}-{\varphi _{7}}^{2}}{r^{2}},}
\\ 
{{f}_{145}^{(+)}{(\varphi )}}={{-f}_{145}^{(-)}{(\varphi )}}={\frac{{\varphi
_{0}}^{2}-{\varphi _{6}}^{2}+{\varphi _{4}}^{2}-{\varphi _{2}}^{2}+{\varphi
_{1}}^{2}-{\varphi _{7}}^{2}+{\varphi _{5}}^{2}-{\varphi _{3}}^{2}}{r^{2}},}
\\ 
{{f}_{176}^{(+)}{(\varphi )}}={{-f}_{176}^{(-)}{(\varphi )}}={\frac{{\varphi
_{0}}^{2}+{\varphi _{6}}^{2}-{\varphi _{4}}^{2}-{\varphi _{2}}^{2}+{\varphi
_{1}}^{2}+{\varphi _{7}}^{2}-{\varphi _{5}}^{2}-{\varphi _{3}}^{2}}{r^{2}},}
\\ 
{{f}_{246}^{(+)}{(\varphi )}}={{-f}_{246}^{(-)}{(\varphi )}}={\frac{{\varphi
_{0}}^{2}+{\varphi _{6}}^{2}+{\varphi _{4}}^{2}+{\varphi _{2}}^{2}-{\varphi
_{1}}^{2}-{\varphi _{7}}^{2}-{\varphi _{5}}^{2}-{\varphi _{3}}^{2}}{{r^{2}}},%
} \\ 
{{f}_{257}^{(+)}{(\varphi )}}={{-f}_{257}^{(-)}{(\varphi )}}={\frac{{\varphi
_{0}}^{2}-{\varphi _{6}}^{2}-{\varphi _{4}}^{2}+{\varphi _{2}}^{2}-{\varphi
_{1}}^{2}+{\varphi _{7}}^{2}+{\varphi _{5}}^{2}-{\varphi _{3}}^{2}}{{r^{2}}},%
} \\ 
{{f}_{347}^{(+)}{(\varphi )}}={{-f}_{347}^{(-)}{(\varphi )}}={\frac{{\varphi
_{0}}^{2}-{\varphi _{6}}^{2}+{\varphi _{4}}^{2}-{\varphi _{2}}^{2}-{\varphi
_{1}}^{2}+{\varphi _{7}}^{2}-{\varphi _{5}}^{2}+{\varphi _{3}}^{2}}{r^{2}},}
\\ 
{{f}_{365}^{(+)}{(\varphi )}}={{-f}_{365}^{(-)}{(\varphi )}}={\frac{{\varphi
_{0}}^{2}+{\varphi _{6}}^{2}-{\varphi _{4}}^{2}-{\varphi _{2}}^{2}-{\varphi
_{1}}^{2}-{\varphi _{7}}^{2}+{\varphi _{5}}^{2}+{\varphi _{3}}^{2}}{r^{2}},}
\end{array}
\]
and the non-standard subset 
\[
\begin{array}{ll}
{{f}_{124}^{(+)}{(\varphi )}}=+2{\frac{{\varphi _{0}}{\varphi _{7}}-{\varphi
_{5}}{\varphi _{2}}+{\varphi _{6}}{\varphi _{1}}+{\varphi _{3}}{\varphi _{4}}%
}{r^{2}},} & {{f}_{125}^{(+)}{(\varphi )}}=-2{\frac{{\varphi _{0}}{\varphi
_{6}}-{\varphi _{3}}{\varphi _{5}}-{\varphi _{1}}{\varphi _{7}}-{\varphi _{2}%
}{\varphi _{4}}}{r^{2}},} \\ 
{{f}_{126}^{(+)}{(\varphi )}}=+2{\frac{{\varphi _{0}}{\varphi _{5}}-{\varphi
_{1}}{\varphi _{4}}+{\varphi _{7}}{\varphi _{2}}+{\varphi _{3}}{\varphi _{6}}%
}{r^{2}},} & {{f}_{127}^{(+)}{(\varphi )}}=-2{\frac{{\varphi _{0}}{\varphi
_{4}}+{\varphi _{6}}{\varphi _{2}}+{\varphi _{1}}{\varphi _{5}}-{\varphi _{3}%
}{\varphi _{7}}}{r^{2}},} \\ 
{{f}_{143}^{(+)}{(\varphi )}}=+2{\frac{{\varphi _{0}}{\varphi _{6}}+{\varphi
_{3}}{\varphi _{5}}+{\varphi _{2}}{\varphi _{4}}-{\varphi _{1}}{\varphi _{7}}%
}{r^{2}},} & {{f}_{146}^{(+)}{(\varphi )}}=-2{\frac{{\varphi _{3}}{\varphi
_{0}}-{\varphi _{4}}{\varphi _{7}}-{\varphi _{1}}{\varphi _{2}}-{\varphi _{5}%
}{\varphi _{6}}}{r^{2}},} \\ 
{{f}_{175}^{(+)}{(\varphi )}}=+2{\frac{{\varphi _{3}}{\varphi _{0}}-{\varphi
_{1}}{\varphi _{2}}+{\varphi _{5}}{\varphi _{6}}+{\varphi _{4}}{\varphi _{7}}%
}{r^{2}},} & {{f}_{247}^{(+)}{(\varphi )}}=-2{\frac{{\varphi _{0}}{\varphi
_{1}}-{\varphi _{7}}{\varphi _{6}}-{\varphi _{4}}{\varphi _{5}}-{\varphi _{3}%
}{\varphi _{2}}}{r^{2}},} \\ 
{{f}_{147}^{(+)}{(\varphi )}}=+2{\frac{{\varphi _{2}}{\varphi _{0}}-{\varphi
_{4}}{\varphi _{6}}+{\varphi _{5}}{\varphi _{7}}+{\varphi _{1}}{\varphi _{3}}%
}{r^{2}},} & {{f}_{243}^{(+)}{(\varphi )}}=-2{\frac{{\varphi _{0}}{\varphi
_{5}}+{\varphi _{1}}{\varphi _{4}}+{\varphi _{7}}{\varphi _{2}}-{\varphi _{3}%
}{\varphi _{6}}}{r^{2}},} \\ 
{{f}_{253}^{(+)}{(\varphi )}}=+2{\frac{{\varphi _{0}}{\varphi _{4}}-{\varphi
_{1}}{\varphi _{5}}+{\varphi _{6}}{\varphi _{2}}+{\varphi _{3}}{\varphi _{7}}%
}{r^{2}},} & {{f}_{173}^{(+)}{(\varphi )}}=-2{\frac{{\varphi _{0}}{\varphi
_{5}}-{\varphi _{7}}{\varphi _{2}}-{\varphi _{3}}{\varphi _{6}}-{\varphi _{1}%
}{\varphi _{4}}}{r^{2}},} \\ 
{{f}_{245}^{(+)}{(\varphi )}}=+2{\frac{{\varphi _{3}}{\varphi _{0}}+{\varphi
_{5}}{\varphi _{6}}-{\varphi _{4}}{\varphi _{7}}+{\varphi _{1}}{\varphi _{2}}%
}{r^{2}},} & {{f}_{256}^{(+)}{(\varphi )}}=+2{\frac{{\varphi _{0}}{\varphi
_{1}}-{\varphi _{3}}{\varphi _{2}}+{\varphi _{7}}{\varphi _{6}}+{\varphi _{4}%
}{\varphi _{5}}}{r^{2}},} \\ 
{{f}_{361}^{(+)}{(\varphi )}}=+2{\frac{{\varphi _{0}}{\varphi _{4}}+{\varphi
_{3}}{\varphi _{7}}+{\varphi _{1}}{\varphi _{5}}-{\varphi _{6}}{\varphi _{2}}%
}{r^{2}},} & {{f}_{362}^{(+)}{(\varphi )}}=-2{\frac{{\varphi _{0}}{\varphi
_{7}}-{\varphi _{3}}{\varphi _{4}}-{\varphi _{5}}{\varphi _{2}}-{\varphi _{6}%
}{\varphi _{1}}}{r^{2}},} \\ 
{{f}_{345}^{(+)}{(\varphi )}}=-2{\frac{{\varphi _{2}}{\varphi _{0}}-{\varphi
_{5}}{\varphi _{7}}-{\varphi _{1}}{\varphi _{3}}-{\varphi _{4}}{\varphi _{6}}%
}{r^{2}},} & {{f}_{346}^{(+)}{(\varphi )}}=+2{\frac{{\varphi _{0}}{\varphi
_{1}}+{\varphi _{3}}{\varphi _{2}}+{\varphi _{7}}{\varphi _{6}}-{\varphi _{4}%
}{\varphi _{5}}}{r^{2}},} \\ 
{{f}_{367}^{(+)}{(\varphi )}}=+2{\frac{{\varphi _{2}}{\varphi _{0}}+{\varphi
_{4}}{\varphi _{6}}+{\varphi _{5}}{\varphi _{7}}-{\varphi _{1}}{\varphi _{3}}%
}{{r^{2}}},} & {{f}_{135}^{(+)}{(\varphi )}}=-2{\frac{{\varphi _{0}}{\varphi
_{7}}-{\varphi _{3}}{\varphi _{4}}+{\varphi _{6}}{\varphi _{1}}+{\varphi _{5}%
}{\varphi _{2}}}{r^{2}},} \\ 
{{f}_{156}^{(+)}{(\varphi )}}=-2{\frac{{\varphi _{2}}{\varphi _{0}}+{\varphi
_{1}}{\varphi _{3}}+{\varphi _{4}}{\varphi _{6}}-{\varphi _{5}}{\varphi _{7}}%
}{r^{2}},} & {{f}_{237}^{(+)}{(\varphi )}}=+2{\frac{{\varphi _{0}}{\varphi
_{6}}-{\varphi _{2}}{\varphi _{4}}+{\varphi _{3}}{\varphi _{5}}+{\varphi _{1}%
}{\varphi _{7}}}{r^{2}},} \\ 
{{f}_{267}^{(+)}{(\varphi )}}=-2{\frac{{\varphi _{3}}{\varphi _{0}}-{\varphi
_{5}}{\varphi _{6}}+{\varphi _{4}}{\varphi _{7}}+{\varphi _{1}}{\varphi _{2}}%
}{r^{2}},} & {{f}_{357}^{(+)}{(\varphi )}}=+2{\frac{{\varphi _{0}}{\varphi
_{1}}+{\varphi _{4}}{\varphi _{5}}-{\varphi _{7}}{\varphi _{6}}+{\varphi _{3}%
}{\varphi _{2}}}{r^{2}},} \\ 
{{f}_{456}^{(+)}{(\varphi )}}=-2{\frac{{\varphi _{0}}{\varphi _{7}}+{\varphi
_{5}}{\varphi _{2}}+{\varphi _{3}}{\varphi _{4}}-{\varphi _{6}}{\varphi _{1}}%
}{r^{2}},} & {{f}_{457}^{(+)}{(\varphi )}}=+2{\frac{{\varphi _{0}}{\varphi
_{6}}+{\varphi _{2}}{\varphi _{4}}+{\varphi _{1}}{\varphi _{7}}-{\varphi _{3}%
}{\varphi _{5}}}{r^{2}},} \\ 
{{f}_{467}^{(+)}{(\varphi )}}=-2{\frac{{\varphi _{0}}{\varphi _{5}}+{\varphi
_{1}}{\varphi _{4}}+{\varphi _{3}}{\varphi _{6}}-{\varphi _{7}}{\varphi _{2}}%
}{r^{2}},} & {{f}_{567}^{(+)}{(\varphi )}}=+2{\frac{{\varphi _{0}}{\varphi
_{4}}-{\varphi _{6}}{\varphi _{2}}-{\varphi _{1}}{\varphi _{5}}-{\varphi _{3}%
}{\varphi _{7}}}{r^{2}}.}
\end{array}
\]
For right actions, the non-standard cocycles are 
\[
\begin{array}{ll}
{{f}_{124}^{(-)}{(\varphi )}}=+2{\frac{{\varphi _{0}}{\varphi _{7}}+{\varphi
_{5}}{\varphi _{2}}-{\varphi _{6}}{\varphi _{1}}-{\varphi _{3}\varphi _{4}}}{%
r^{2}},} & {{f}_{125}^{(-)}{(\varphi )}}=-2{\frac{{\varphi _{0}}{\varphi _{6}%
}+{\varphi _{3}}{\varphi _{5}}+{\varphi _{1}}{\varphi _{7}}+{\varphi _{2}}{%
\varphi _{4}}}{r^{2}},} \\ 
{{f}_{126}^{(-)}{(\varphi )}}=+2{\frac{{\varphi _{0}}{\varphi _{5}}+{\varphi
_{1}}{\varphi _{4}}-{\varphi _{7}}{\varphi _{2}}-{\varphi _{3}\varphi _{6}}}{%
r^{2}},} & {{f}_{127}^{(-)}{(\varphi )}}=-2{\frac{{\varphi _{0}}{\varphi _{4}%
}-{\varphi _{6}}{\varphi _{2}}-{\varphi _{1}}{\varphi _{5}}+{\varphi _{3}}{%
\varphi _{7}}}{r^{2}},} \\ 
{{f}_{143}^{(-)}{(\varphi )}}=+2{\frac{{\varphi _{0}}{\varphi _{6}}-{\varphi
_{3}}{\varphi _{5}}-{\varphi _{2}}{\varphi _{4}}+{\varphi _{1}\varphi _{7}}}{%
r^{2}},} & {{f}_{146}^{(-)}{(\varphi )}}=-2{\frac{{\varphi _{3}}{\varphi _{0}%
}+{\varphi _{4}}{\varphi _{7}}+{\varphi _{1}}{\varphi _{2}}+{\varphi _{5}}{%
\varphi _{6}}}{r^{2}},} \\ 
{{f}_{175}^{(-)}{(\varphi )}}=+2{\frac{{\varphi _{3}}{\varphi _{0}}+{\varphi
_{1}}{\varphi _{2}}-{\varphi _{5}}{\varphi _{6}}-{\varphi _{4}\varphi _{7}}}{%
r^{2}},} & {{f}_{247}^{(-)}{(\varphi )}}=-2{\frac{{\varphi _{0}}{\varphi _{1}%
}+{\varphi _{7}}{\varphi _{6}}+{\varphi _{4}}{\varphi _{5}}+{\varphi _{3}}{%
\varphi _{2}}}{r^{2}},} \\ 
{{f}_{147}^{(-)}{(\varphi )}}=+2{\frac{{\varphi _{2}}{\varphi _{0}}+{\varphi
_{4}}{\varphi _{6}}-{\varphi _{5}}{\varphi _{7}}-{\varphi _{1}\varphi _{3}}}{%
r^{2}},} & {{f}_{243}^{(-)}{(\varphi )}}=-2{\frac{{\varphi _{0}}{\varphi _{5}%
}-{\varphi _{1}}{\varphi _{4}}-{\varphi _{7}}{\varphi _{2}}+{\varphi _{3}}{%
\varphi _{6}}}{r^{2}},} \\ 
{{f}_{253}^{(-)}{(\varphi )}}=+2{\frac{{\varphi _{0}}{\varphi _{4}}+{\varphi
_{1}}{\varphi _{5}}-{\varphi _{6}}{\varphi _{2}}-{\varphi _{3}\varphi _{7}}}{%
r^{2}},} & {{f}_{173}^{(-)}{(\varphi )}}=-2{\frac{{\varphi _{0}}{\varphi _{5}%
}+{\varphi _{7}}{\varphi _{2}}+{\varphi _{3}}{\varphi _{6}}+{\varphi _{1}}{%
\varphi _{4}}}{r^{2}},} \\ 
{{f}_{245}^{(-)}{(\varphi )}}=+2{\frac{{\varphi _{3}}{\varphi _{0}}-{\varphi
_{5}}{\varphi _{6}}+{\varphi _{4}}{\varphi _{7}}-{\varphi _{1}\varphi _{2}}}{%
r^{2}},} & {{f}_{256}^{(-)}{(\varphi )}}=+2{\frac{{\varphi _{0}}{\varphi _{1}%
}+{\varphi _{3}}{\varphi _{2}}-{\varphi _{7}}{\varphi _{6}}-{\varphi _{4}}{%
\varphi _{5}}}{r^{2}},} \\ 
{{f}_{361}^{(-)}{(\varphi )}}=+2{\frac{{\varphi _{0}}{\varphi _{4}}-{\varphi
_{3}}{\varphi _{7}}-{\varphi _{1}}{\varphi _{5}}+{\varphi _{6}\varphi _{2}}}{%
r^{2}},} & {{f}_{362}^{(-)}{(\varphi )}}=-2{\frac{{\varphi _{0}}{\varphi _{7}%
}+{\varphi _{3}}{\varphi _{4}}+{\varphi _{5}}{\varphi _{2}}+{\varphi _{6}}{%
\varphi _{1}}}{r^{2}},} \\ 
{{f}_{345}^{(-)}{(\varphi )}}=-2{\frac{{\varphi _{2}}{\varphi _{0}}+{\varphi
_{5}}{\varphi _{7}}+{\varphi _{1}}{\varphi _{3}}+{\varphi _{4}\varphi _{6}}}{%
r^{2}},} & {{f}_{346}^{(-)}{(\varphi )}}=+2{\frac{{\varphi _{0}}{\varphi _{1}%
}-{\varphi _{3}}{\varphi _{2}}-{\varphi _{7}}{\varphi _{6}}+{\varphi _{4}}{%
\varphi _{5}}}{r^{2}},} \\ 
{{f}_{367}^{(-)}{(\varphi )}}=+2{\frac{{\varphi _{2}}{\varphi _{0}}-{\varphi
_{4}}{\varphi _{6}}-{\varphi _{5}}{\varphi _{7}}+{\varphi _{1}\varphi _{3}}}{%
{r^{2}}},} & {{f}_{135}^{(-)}{(\varphi )}}=-2{\frac{{\varphi _{0}}{\varphi
_{7}}+{\varphi _{3}}{\varphi _{4}}-{\varphi _{6}}{\varphi _{1}}-{\varphi _{5}%
}{\varphi _{2}}}{r^{2}},} \\ 
{{f}_{156}^{(-)}{(\varphi )}}=-2{\frac{{\varphi _{2}}{\varphi _{0}}-{\varphi
_{1}}{\varphi _{3}}-{\varphi _{4}}{\varphi _{6}}+{\varphi _{5}\varphi _{7}}}{%
r^{2}},} & {{f}_{237}^{(-)}{(\varphi )}}=+2{\frac{{\varphi _{0}}{\varphi _{6}%
}+{\varphi _{2}}{\varphi _{4}}-{\varphi _{3}}{\varphi _{5}}-{\varphi _{1}}{%
\varphi _{7}}}{r^{2}},} \\ 
{{f}_{267}^{(-)}{(\varphi )}}=-2{\frac{{\varphi _{3}}{\varphi _{0}}+{\varphi
_{5}}{\varphi _{6}}-{\varphi _{4}}{\varphi _{7}}-{\varphi _{1}\varphi _{2}}}{%
r^{2}},} & {{f}_{357}^{(-)}{(\varphi )}}=+2{\frac{{\varphi _{0}}{\varphi _{1}%
}-{\varphi _{4}}{\varphi _{5}}+{\varphi _{7}}{\varphi _{6}}-{\varphi _{3}}{%
\varphi _{2}}}{r^{2}},} \\ 
{{f}_{456}^{(-)}{(\varphi )}}=-2{\frac{{\varphi _{0}}{\varphi _{7}}-{\varphi
_{5}}{\varphi _{2}}-{\varphi _{3}}{\varphi _{4}}+{\varphi _{6}\varphi _{1}}}{%
r^{2}},} & {{f}_{457}^{(-)}{(\varphi )}}=+2{\frac{{\varphi _{0}}{\varphi _{6}%
}-{\varphi _{2}}{\varphi _{4}}-{\varphi _{1}}{\varphi _{7}}+{\varphi _{3}}{%
\varphi _{5}}}{r^{2}},} \\ 
{{f}_{467}^{(-)}{(\varphi )}}=-2{\frac{{\varphi _{0}}{\varphi _{5}}-{\varphi
_{1}}{\varphi _{4}}-{\varphi _{3}}{\varphi _{6}}+{\varphi _{7}\varphi _{2}}}{%
r^{2}},} & {{f}_{567}^{(-)}{(\varphi )}}=+2{\frac{{\varphi _{0}}{\varphi _{4}%
}+{\varphi _{6}}{\varphi _{2}}+{\varphi _{1}}{\varphi _{5}}+{\varphi _{3}}{%
\varphi _{7}}}{r^{2}}.}
\end{array}
\]


\newpage

\begin{thebibliography}{99}
\bibitem{s1}  P. Jordan, J Von Neumann and E. Wigner, Ann. of Math. 35
(1934) 29.

\bibitem{s2}  R. E. Behrends, J. Dreitlein, C. Fronsdal and B.W. Lee, Rev.
Mod. Phys. 34 (1962) 1, D.R. Speiser and J. Tarski J. Math. Phys. 4 (1963)
588, J. Souriau and D. Kastler, C. R. Acad. Sci. Paris 250 (1960) 2807, A.
Gamba, J. Math. Phys. 8 (1967) 775, F. Gursey, Ann. of Phys. 12 (1961) 91,
D. Horn and Y.Ne'eman, Nouvo Cimento 31 (1964) 879, M. Gourdin, Nouvo
Cimento 30 (1963) 587, G. Cocho, Phys. Rev. B137 (1965) 1255, P. Gueret, J.
P. Vigier and W. Tait, Nouvo Cimento A17 (1973) 663.

\bibitem{s3}  M. Gunaydin and F. Gursey, J. Math. Phys. 14 (1973) 1651, M.
Gunaydin and F. Gursey, Phys. Rev. D9 (1974) 3387.

\bibitem{s4}  F. Gursey,``Exceptional Groups And Elementary Particles.
(Talk),'' \textit{In Nijmegen 1975, Proceedings, Group Theoretical Methods
In Physics, Berlin 1976, 225-233}.

\bibitem{s5}  F. Gursey, P. Ramond and P. Sikivie, Phys. Lett. B60 (1976)
177, P. Sikivie and F. Gursey, Phys. Rev. D16 (1977) 816.

\bibitem{s6}  M.~.F.~Atiyah, V.~G.~Drinfield, N.~J.~Hitchin, and
Yu.~I.~Manin, Phys.~Lett. 65A (1978) 185.

\bibitem{s7}  T.~Kugo and P.~Townsend, Nucl. Phys. \textbf{B221} (1983) 357,
K .~W.~Chung~and~A.~Sudbery, Phys.~Lett. B198 (1987) 161, J.~M.~Evans,
Nucl.~Phys. B298 (1988) 92, J.~M.~Evans, Nucl.~Phys. \textbf{B310} (1988) 44
, N.~Berkovits, Phys. Lett. B318 (1993) 104.

\bibitem{s8}  M. J. Duff, B. E. Nilsson and C. N. Pope, Phys. Rept. 130
(1986) 1.

\bibitem{s9}  F. Englert, Phys. Lett. Phys. Lett. B119 (1982) 339.

\bibitem{s10}  M.~Rooman, Nucl. Phys. \textbf{B236} (1984) 501.

\bibitem{s11}  F. Gursey and C. Tze, Phys. Lett. B127 (1983) 191, T. Dereli,
M. Panahimoghaddam, A. Sudbery and R. W. Tucker, Phys. Lett. B126 (1983) 33,
B. de Wit and H. Nicolai, Nucl. Phys. B231 (1984) 506.

\bibitem{s12}  M.~J.~Duff, R.~R.~Khuri and J.~X.~Lu, Phys. Rep. \textbf{259}
(1995) 213, J.~A.~Harvey and A.~Strominger, Phys. Rev. Lett. \textbf{66}
(1991) 549, M.~J.~Duff, J.~M.~Evans, R.~R.~Khuri, J.~X.~Lu, R.~Minasian,
``The Octonionic Membrane'', hep-th/9706124, B.~S.~Acharya, M.~O'Loughlin,
and B.~Spence, Nucl. Phys. \textbf{503B} (1997), 657, L.~Baulieu, H.~Kanno,
and I.~M.~Singer, ``Cohomological Yang-Mills theory in eight dimensions'',
hep-th/9705127, G.~Gibbons, G.~Papadopoulos, and K.~Stelle, ``HKT and OKT
geometries on soliton black hole moduli spaces'', hep-th/9706207,
C.~M.~Hull, ``Higher dimensional Yang-Mills theories and topological
terms'', hep-th/9710165, J.~M.~Figueroa-O'Farrill, ``Gauge theory and the
Division Algebras'', hep-th/9710168.

\bibitem{s13}  M.~Sohnius, Z. Phys. \textbf{C18} (1983) 229.

\bibitem{s14}  F.~Englert, A.~Servin, W.~Troost, A.~Van~Proeyen and
Ph.~Spindel, J.~Math.~Phys. \textbf{29} (1988) 281.

\bibitem{s15}  N. Berkovits, Phys.~Lett. \textbf{241B} (1990) 497,
N.~Berkovits, Nucl.~Phys. \textbf{358B} (1991) 169.

\bibitem{s16}  M. Cederwall and C.~R.Preitschopf, Comm.~Math.~Phys. \textbf{%
167} (1995) 373.

\bibitem{s17}  L.~Brink, M.~Cederwall and C.~R.~Preitschopf, Phys.~Lett. 
\textbf{311B} (1993) 76.

\bibitem{s18}  J. A. H. Samtleben, Nucl. Phys. \textbf{453B} (95) 429.

\bibitem{s19}  R. D. Schafer, An introduction to non-associative algebrae
(Academic Press, New York, 1966).

\bibitem{s20}  S.~De~Leo~and~K.~Abdel-Khalek, J. Math. Phys., 38 (1997) 582.
K. Abdel-Khalek, Int. J. Mod. Phys. A13 (1998) 569.

\bibitem{s21}  P.~Rotelli, Mod.~Phys.~Lett. A., \textbf{4} (1989) 933.

\bibitem{s22}  E. Cartan and J. A. Schouten, Proc. Kon. Akad. Wet.
Amesterdam 29 (1926) 803, 933. E. Cartan, J. Math. Pures et Appl. 6 (1927)
1. J.~A.~Wolf, J.~Diff.~Geom. 6 (1972) 317.

\bibitem{s23}  M.~Gunaydin and S.V.~Ketov, Nucl. Phys. B467 (1996) 215,
M.~Gunaydin and H.~Nicolai, Phys.~Lett. 351B (1995) 169, B. de Wit and H.
Nicolai, Nucl. Phys. B231 (1984) 506.

\bibitem{s24}  C.A.~Manogue and J.~Schray, J. Math. Phys. \textbf{34} (1993)
3746 hep-th/9302044, J.~Schray and C.A.~Manogue, hep-th/9407179.

\bibitem{s25}  E. Cartan, The theory of spinors, (Cambridge, MA:MIT press,
1966).

\bibitem{s26}  J.~Lukierski and P.~Minnaert, Phys. Lett. \textbf{129B}
(1983) 392.
\end{thebibliography}
\end{document}